\documentclass[aps,pra,superscriptaddress,twocolumn,showpacs]{revtex4}
\usepackage{amsfonts}
\usepackage{amsmath}
\usepackage{amssymb}
\usepackage{graphicx}

\begin{document}

\title{Stationary states and quantum quench dynamics of Bose-Einstein
condensates in a double-well potential}
\author{Linghua Wen}
\email{linghuawen@ysu.edu.cn}
\affiliation{College of Science, Yanshan University, Qinhuangdao 066004, China}
\author{Qizhong Zhu}
\affiliation{Department of Physics and Center of Theoretical and Computational Physics,
The University of Hong Kong, Hong Kong, China}
\author{Tianfu Xu}
\affiliation{College of Science, Yanshan University, Qinhuangdao 066004, China}
\author{Xili Jing}
\affiliation{College of Science, Yanshan University, Qinhuangdao 066004, China}
\author{Chengshi Liu}
\affiliation{College of Science, Yanshan University, Qinhuangdao 066004, China}
\date{\today }

\begin{abstract}
We consider the properties of stationary states and the dynamics of
Bose-Einstein condensates (BECs) in a double-well (DW) potential with pair
tunneling by using a full quantum-mechanical treatment. Furthermore, we
study the quantum quench dynamics of the DW system subjected to a sudden
change of the Peierls phase. It is shown that strong pair tunneling
evidently\ influences the energy spectrum structure of the stationary
states. For relatively weak repulsive interatomic interactions, the dynamics
of the DW system with a maximal initial population difference evolves from
Josephson oscillations to quantum self-trapping as one increases the pair
tunneling strength, while for large repulsion the strong pair tunneling
inhibits the quantum self-trapping. In the case of attractive interatomic
interactions, strong pair tunneling tends to destroy the Josephson
oscillations and quantum self-trapping, and the system eventually enters a
symmetric regime of zero population difference. Finally, the effect of the
Peierls phase on the quantum quench dynamics of the system is analyzed and
discussed. These new features are remarkably different from the usual
dynamical behaviors of a BEC in a DW potential.
\end{abstract}

\maketitle

\section{Introduction}

Stationary state properties and dynamics of ultracold atoms in a double-well
(DW) potential have attracted considerable attention over the past few
decades \cite{Leggett,Ueda}. As a powerful building block model, the double
well plays a key role in revealing numerous interesting phenomena of quantum
many-body and few-body systems due to the experimental accessibility and
precise controllability. Actually, many intriguing properties have been
predicted theoretically and some observed experimentally in\ Bose-Einstein
condensates (BECs) or ultracold fermionic atoms in a DW potential, ranging
from the Josephson effect and quantum self-trapping \cite%
{Smerzi,Milburn,Pu,Albiez,Ananikian,Wen1,Shchesnovich,Adhikari,JuliaDiaz1,Gillet,Gati,Edmonds}%
, fragmentation of a BEC \cite{Spekkens,JuliaDiaz2}, entangled clusters \cite%
{YWu} and NOON-like states \cite{Carr} of condensed bosons, quantum chaos
\cite{Mahmud}, hidden vortices \cite{Wen2,Wen3} and vortex superpositions
\cite{Garcia}, and spin correlation \cite{Carvalho} to two-particle analog
of a charge-density-wave state \cite{Murmann}, etc. Of particular interest
is the quantum tunneling dynamics in a DW BEC because it represents one of
the most surprising and paradigmatic effects in quantum mechanics. Although
tunneling is a fundamental phenomenon in wave dynamics (for instance, the
dynamics of atomic matter waves and the classical dynamics of optical waves)
\cite{Malomed}, the quantum tunneling of ultracold bosonic atoms through a
potential barrier between two wells provides a direct manifestation of
quantum phase coherence. Recently, a correlated quantum tunneling was
observed in Ref. \cite{Folling} and a relevant theoretical analysis was
presented by means of two-body dynamics \cite{Zollner}. Further study
demonstrates that correlated tunneling is principally resulted from pair
tunneling, which involves the superexchange interactions between particles
on neighboring wells \cite{Liang}. Most recently, spontaneous symmetry
breaking and effective ground state in a DW BEC with pair tunneling are
discussed in Ref. \cite{Zhu}. These theoretical investigations \cite%
{Zollner,Liang,Zhu}\ concerning to the pair tunneling are confined to the
regime of repulsive interatomic interactions. In addition, the analyses in
Refs. \cite{Zollner,Liang}\ are illustrated with a two-atom or few-body
system. However, the usual realistic ultracold atomic gases in experiments
should be many-body systems. In fact, strong many-body effect may basically
alter the tunneling configuration of the DW system.

On the other hand, the research on neutral atoms in synthetic gauge fields
has become an active new subject in the filed of cold atom physics \cite%
{Dalibard}. The artificial gauge potentials constitute novel tools for
exploring the properties of ultracold atoms, which usually occur in the
field of condensed matter physics. Recently, Spielman's group experimentally
realized a Peierls substitution for ultracold neutral atoms in an artificial
lattice potential \cite{Jimenez}, where the Peierls phase can be controlled
precisely. The effective vector gauge potential is characterized by a
complex tunneling parameter $J=\left\vert J\right\vert e^{i\theta }$, where $%
\theta $ is the so-called Peierls phase gained by an atom tunneling from
site $j+1$ to $j$. In addition, the Peierls phase may also be created in a
shaken optical lattice \cite{Struck,Zheng} or a spin-dependent optical
lattice \cite{Greschner}. In these studies, the equilibrium states of the
many-body system can be well described. However, little is known about the
quantum quench dynamics of a system far from equilibrium, especially for the
quantum quenches of the Hamiltonian caused by a sudden change of the Peierls
phase. From a fundamental viewpoint, the quantum quench dynamics of cold
atom systems far from equilibrium is very interesting because it can reveal
rich properties of the many-body system beyond ground state. For instance,
\textit{Zitterbewegung} oscillation and dynamical topological phases have
been observed most recently in a quenched spin-orbit-coupled (SOC) BEC \cite%
{Qu} and a quenched SOC degenerate Fermi gas \cite{Dong}, respectively.

In this work we investigate the stationary state properties of a BEC in a DW
potential with pair tunneling and the quantum dynamics of the DW system with
a maximal initial population difference as well as the quench dynamics of
the system by abruptly changing the Peierls phase. Although the problems of
bosons in a DW potential can be solved under certain conditions by several
semi-classical analytical approximations, such as mean-field theory \cite%
{Smerzi,Milburn,Ananikian}, variational approach to many-body ground state
\cite{Spekkens}, phase-space analysis of many-body bosons \cite%
{JuliaDiaz1,Mahmud} and WKB approach \cite{Shchesnovich}, these
semi-classical treatments are usually too complicated and redundant in
understanding the physical essence of the problems. In particular, the
correlated tunneling observed in Ref. \cite{Folling} can not be described by
the standard Bose-Hubbard (BH) Hamiltonian \cite{Jaksch} based on the
hard-core interaction under the semi-classical approximations. Here we adopt
a tractable full quantum-mechanical scheme to tackle the present problem.
The main features are summarized as follows. First, strong pair tunneling
can make the eigenstates of the system evolves from delocalized states to
Schr\"{o}dinger cat-like states and quantum self-trapping states and vice
versa, depending on the many-body interactions. Second, for the case of
relatively weak repulsive interactions strong pair tunneling sustains a
quantum self-trapping, but for the case of large repulsive interactions the
strong pair tunneling tends to destroy the self-trapping. On the other hand,
for attractive interatomic interactions the strong pair tunneling tends to
eliminate the Josephson oscillation and quantum self-trapping irrespective
of the interaction strength. Finally, once the DW system is quenched, the
population difference shows evident oscillation behaviors for most cases.
The maximum oscillation amplitude is directly relevant to the value of the
Peierls phase. These interesting features are markedly different from the
usual stationary state structures and dynamical behaviors of a BEC in a DW
potential, which allows to be observed and tested under current experimental
technique conditions.

The paper is organized as follows. In Sec. II, the model Hamiltonian is
introduced. The general properties of the stationary states of the system
are described in Sec. III. The dynamics of the system with a maximal initial
population difference is investigated in Sec. IV. The quantum quench
dynamics of the DW system is discussed in Sec. V. Conclusions are outlined
in Sec. VI.

\section{Model}

We consider a system of $N$ ultracold bosonic atoms in a DW potential. At
low temperatures the second-quantized Hamiltonian of the system beyond the
onsite approximation can be written as \cite{Liang,Zhu,Jimenez,Bader,Dutta}%
\begin{eqnarray}
\hat{H} &=&-\left[ J-U_{3}\left( \hat{n}_{1}+\hat{n}_{2}-1\right) \right]
\left( e^{i\theta }\hat{a}_{1}^{\dag }\hat{a}_{2}+e^{-i\theta }\hat{a}%
_{2}^{\dag }\hat{a}_{1}\right)  \notag \\
&&+\frac{U_{0}}{2}\left( \hat{a}_{1}^{\dagger }\hat{a}_{1}^{\dagger }\hat{a}%
_{1}\hat{a}_{1}+\hat{a}_{2}^{\dag }\hat{a}_{2}^{\dag }\hat{a}_{2}\hat{a}%
_{2}\right) +\left( U_{1}+U_{2}\right) \hat{n}_{1}\hat{n}_{2}  \notag \\
&&+\frac{U_{2}}{2}\left( e^{2i\theta }\hat{a}_{1}^{\dag }\hat{a}_{1}^{\dag }%
\hat{a}_{2}\hat{a}_{2}+e^{-2i\theta }\hat{a}_{2}^{\dag }\hat{a}_{2}^{\dag }%
\hat{a}_{1}\hat{a}_{1}\right)  \notag \\
&&+\mu \left( \hat{n}_{1}-\hat{n}_{2}\right) ,  \label{StationaryHamiltonian}
\end{eqnarray}%
where $\hat{n}_{j}=\hat{a}_{j}^{\dag }\hat{a}_{j}$ and $\hat{a}_{j}$ ($\hat{a%
}_{j}^{\dag }$) is the bosonic annihilation (creation) operator for well $j$%
. $J$ is the hopping amplitude of single-particle tunneling, $U_{0}$ is the
on-site two-body interaction strength, $U_{1}$ is the inter-well particle
interaction, $U_{2}$ describes the atom-pair tunneling (pair tunneling), and
$U_{3}$ denotes the density-dependent tunneling \cite{Liang,Zhu,Dutta}. $%
\theta $ represents a tunable Peierls phase \cite{Jimenez}, and the tilt
parameter $\mu $ denotes the difference of local chemical potentials caused
for instance by a mismatch between the two wells. In general, $U_{1}\approx
U_{2}$, and the coupling constants $U_{1}$, $U_{2}$, and $U_{3}$ are smaller
compared with $U_{0}$. Throughout this paper we assume $U_{1}=U_{2}$ for the
DW system and take $U_{3}\approx 2.5U_{2}$, which is consistent with the
choice of the corresponding parameters in Ref. \cite{Folling,Liang}. When $%
\theta =0$ and the terms involving $U_{1}$, $U_{2}$, and $U_{3}$ are
neglected, equation (\ref{StationaryHamiltonian}) becomes the standard BH
Hamiltonian for a DW potential \cite{Jaksch,Dutta}.

By introducing the notations $\hat{H}_{r}=\hat{H}/NJ$, $\widetilde{U}%
_{0}=NU_{0}/2J$, $\widetilde{U}_{2}=NU_{2}/2J$, $\widetilde{U}_{3}=NU_{3}/2J$%
, and $\widetilde{\mu }=\mu /J$, we obtain the rescaled Hamiltonian in
reduced units%
\begin{eqnarray}
\hat{H}_{r} &=&-\left[ \frac{1}{N}-\frac{2U_{3}}{N^{2}}\left( \hat{n}_{1}+%
\hat{n}_{2}-1\right) \right] \left( e^{i\theta }\hat{a}_{1}^{\dag }\hat{a}%
_{2}+e^{-i\theta }\hat{a}_{2}^{\dag }\hat{a}_{1}\right)  \notag \\
&&+\frac{U_{0}}{N^{2}}\left( \hat{a}_{1}^{\dagger }\hat{a}_{1}^{\dagger }%
\hat{a}_{1}\hat{a}_{1}+\hat{a}_{2}^{\dag }\hat{a}_{2}^{\dag }\hat{a}_{2}\hat{%
a}_{2}\right) +\frac{4U_{2}}{N^{2}}\hat{n}_{1}\hat{n}_{2}  \notag \\
&&+\frac{U_{2}}{N^{2}}\left( e^{2i\theta }\hat{a}_{1}^{\dag }\hat{a}%
_{1}^{\dag }\hat{a}_{2}\hat{a}_{2}+e^{-2i\theta }\hat{a}_{2}^{\dag }\hat{a}%
_{2}^{\dag }\hat{a}_{1}\hat{a}_{1}\right)  \notag \\
&&+\frac{\mu }{N}\left( \hat{n}_{1}-\hat{n}_{2}\right) ,
\label{ReducedHamiltonian}
\end{eqnarray}%
where the tilde\ is omitted for simplicity. The most general $N$-body state
vector is a linear superposition of Fock states%
\begin{equation}
\left\vert \psi \right\rangle =\sum\limits_{k=1}^{N+1}c_{k}\left\vert
k-1,N-k+1\right\rangle ,  \label{StateVector}
\end{equation}%
with $k-1$ being the occupation number corresponding to the left well and
\begin{equation}
\left\vert k-1,N-k+1\right\rangle =\frac{e^{i(k-1)\theta }\left( \hat{a}%
_{1}^{\dagger }\right) ^{k-1}\left( \hat{a}_{2}^{\dag }\right) ^{N-k+1}}{%
\sqrt{\left( k-1\right) !\left( N-k+1\right) !}}\left\vert vac\right\rangle .
\label{FockState}
\end{equation}%
The number of atoms in the $j$th well is $N_{j}=\left\langle \psi
\right\vert \hat{a}_{j}^{\dag }\hat{a}_{j}\left\vert \psi \right\rangle $, $%
N=N_{1}+N_{2}$ is the total number of atoms, and the population difference
is defined as $z=\left( N_{1}-N_{2}\right) /N$. The generic properties of
the stationary states of the system can be obtained by solving the
eigenvalue equation of the Hamiltonian (\ref{ReducedHamiltonian}).

The dynamics of the system is described by the time-dependent Schr\"{o}%
dinger equation%
\begin{equation}
\frac{i}{N}\frac{\partial }{\partial \tau }\left\vert \Psi \right\rangle =%
\hat{H}_{r}\left\vert \Psi \right\rangle ,  \label{DynamicEquation}
\end{equation}%
where the time is measured in tunneling time units: $\tau =t/T$ with $%
T=\hbar /J$. The time-dependent solution of equation (\ref{DynamicEquation})
can be expanded in the Fock basis%
\begin{equation}
\left\vert \Psi (\tau )\right\rangle =\sum\limits_{k=1}^{N+1}C_{k}(\tau
)\left\vert k-1,N-k+1\right\rangle .  \label{TimeDependentStateVector}
\end{equation}%
Combining equations (\ref{DynamicEquation}) and (\ref%
{TimeDependentStateVector}), we can obtain the coupled equations for the
time evolution of the coefficients $C_{k}$%
\begin{eqnarray}
\frac{i}{N}\frac{d}{d\tau }C_{1} &=&\left[ U_{0}(1-\frac{1}{N})-\mu \right]
C_{1}  \notag \\
&&+\left[ \frac{2U_{3}(N-1)}{N^{3/2}}-\frac{1}{\sqrt{N}}\right] e^{-i\theta
}C_{2}  \notag \\
&&+\frac{U_{2}\sqrt{2N(N-1)}}{N^{2}}e^{-2i\theta }C_{3},  \label{C1Evolution}
\end{eqnarray}%
\begin{eqnarray}
\frac{i}{N}\frac{d}{d\tau }C_{2} &=&\frac{U_{0}(N-1)(N-2)+4U_{2}(N-1)}{N^{2}}%
C_{2}  \notag \\
&&+\mu \left( \frac{2}{N}-1\right) C_{2}  \notag \\
&&+\frac{2U_{3}(N-1)-N}{N^{3/2}}e^{i\theta }C_{1}  \notag \\
&&+\frac{\left[ 2U_{3}(N-1)-N\right] \sqrt{2(N-1)}}{N^{2}}e^{-i\theta }C_{3}
\notag \\
&&+\frac{U_{2}\sqrt{6(N-1)(N-2)}}{N^{2}}e^{-2i\theta }C_{4},
\label{C2Evolution}
\end{eqnarray}%
\begin{eqnarray}
\frac{i}{N}\frac{d}{d\tau }C_{k} &=&a_{k}C_{k}+b_{k-1}e^{i\theta
}C_{k-1}+b_{k+1}e^{-i\theta }C_{k+1}  \notag \\
&&+d_{k-2}e^{2i\theta }C_{k-2}+d_{k+2}e^{-2i\theta }C_{k+2},\text{ }  \notag
\\
&&\left( k=3,4,...,N-1\right) ,  \label{CkEvolution}
\end{eqnarray}%
\begin{eqnarray}
\frac{i}{N}\frac{d}{d\tau }C_{N} &=&\frac{U_{0}(N-1)(N-2)+4U_{2}(N-1)}{N^{2}}%
C_{N}  \notag \\
&&+\mu \left( 1-\frac{2}{N}\right) C_{N}  \notag \\
&&+\frac{\left[ 2U_{3}(N-1)-N\right] \sqrt{2(N-1)}}{N^{2}}e^{i\theta }C_{N-1}
\notag \\
&&+\frac{2U_{3}(N-1)-N}{N^{3/2}}e^{-i\theta }C_{N+1}  \notag \\
&&+\frac{U_{2}\sqrt{6(N-1)(N-2)}}{N^{2}}e^{2i\theta }C_{N-2},
\label{CNEvolution}
\end{eqnarray}%
\begin{eqnarray}
\frac{i}{N}\frac{d}{d\tau }C_{N+1} &=&\left[ U_{0}\left( 1-\frac{1}{N}%
\right) +\mu \right] C_{N+1}  \notag \\
&&+\frac{U_{2}\sqrt{2N(N-1)}}{N^{2}}e^{2i\theta }C_{N-1}  \notag \\
&&+\frac{2U_{3}(N-1)-N}{N^{3/2}}e^{i\theta }C_{N},  \label{CN1Evolution}
\end{eqnarray}%
where $%
a_{k}=(U_{0}/N^{2})[(k-1)(k-2)+(N-k)(N-k+1)]+(4U_{2}/N^{2})(k-1)(N-k+1)+(\mu
/N)\left[ 2(k-1)-N\right] $, $b_{k-1}=[2U_{3}(N-1)/N^{2}-1/N]\sqrt{%
(k-1)(N-k+2)}$, $b_{k+1}=[2U_{3}(N-1)/N^{2}-1/N]\sqrt{k(N-k+1)}$, $%
d_{k-2}=(U_{2}/N^{2})\sqrt{(k-1)(k-2)(N-k+2)(N-k+3)}$, and $%
d_{k+2}=(U_{2}/N^{2})\sqrt{k(k+1)(N-k)(N-k+1)}$. The time-dependent
population difference between the left and right wells is given by%
\begin{equation}
z(\tau )=\frac{\sum\limits_{k=1}^{N+1}(2k-N-2)\left\vert C_{k}(\tau
)\right\vert ^{2}}{N},  \label{PopulationDifference}
\end{equation}%
where the coefficients are normalized as $\sum\limits_{k=1}^{N+1}\left\vert
C_{k}(\tau )\right\vert =1$.

In the following we systematically investigate the general properties of the
stationary states of the DW system with pair tunneling both for repulsive
and attractive interatomic interactions. Moreover, we discuss the dynamics
of the system with a maximal initial population imbalance. Finally, we
analyze the quantum quench dynamics of the system by suddenly changing the
value of the Peierls phase at a certain moment.

\section{Properties of the stationary states of the DW system with pair
tunneling}

We first consider the stationary state properties of the DW system. The
eigenstates can be obtained through the exact diagonalization \cite%
{JuliaDiaz1} of the Hamiltonian (\ref{ReducedHamiltonian}) in the $(N+1)$%
-dimensional space spanned by the Fock basis: $\left\{ \left\vert
0,N\right\rangle ,\left\vert 1,N-1\right\rangle ,...,\left\vert
N-1,1\right\rangle ,\left\vert N,0\right\rangle \right\} $. In figure 1, we
show the density profiles $\left\vert c_{k}^{s}\right\vert ^{2}$ of the
eigenvector of the Hamiltonian (\ref{ReducedHamiltonian}) for various
parameters, where $N=50$, $\theta =0$, and $s=1,2,...,N+1$ denotes the
ordinal number of the eigenvector ordered by increasing eigenvalue. For the
case of $U_{0}=0$, $U_{2}=0$, and $\mu =0$ (i.e., there is no interatomic
interaction and the double well is symmetric), the density profile of each
eigenstate displays excellent left-right symmetry [figure 1(a1)], where the
eigenstates are non-degenerate. With the increasing of the interatomic
interaction, the high-lying excited states gradually form degenerate pairs
according to the up-down order. The degeneracy of each pair in the excited
states can be eliminated by artificially introducing a small tilt of the
double well. We show that for a repulsive interatomic interaction $U_{0}$
above a critical value the left-right symmetry of the high-lying states is
broken such that the density profile develops an evident population
imbalance and occupies mainly one region of the Fock space. Consequently,
the highest excited state spontaneously acquires a large population
imbalance, which means the highest-lying state evolves from a delocalized
state to a Schr\"{o}dinger cat-like state and eventually to a quantum
self-trapping state due to the symmetry breaking. The above analysis can
explain why there is a large dropwell of the density profiles in figure
1(a2). When the repulsive interaction $U_{0}$ is sufficiently strong, the
density profiles become almost two cross line segments [figure 1(a3)]. In
this situation, most energy eigenstates are pair quasi-degenerate and
non-extended (cat-like or highly localized). Physically, this point can be
understood because the sufficiently strong repulsive interaction is
equivalent to a very high central barrier of the DW potential, where the
particle tunneling between the two wells is highly suppressed and the
coherence between the two sites is lost. For the case of ground state, a
well-known paradigm is the Mott insulator phase with a commensurating
filling in a optical lattice \cite{Jaksch,Greiner}. Note that the case of
attractive interactions is similar to that of the repulsive interactions,
but the role played by the highest-lying excited state is now replaced by
the ground state of the Hamiltonian (\ref{ReducedHamiltonian}) and the
low-lying eigenstates exhibit two-fold degeneracy [see figure 3(b)].

\begin{figure}[tbp]
\centerline{\includegraphics*[width=8.5cm]{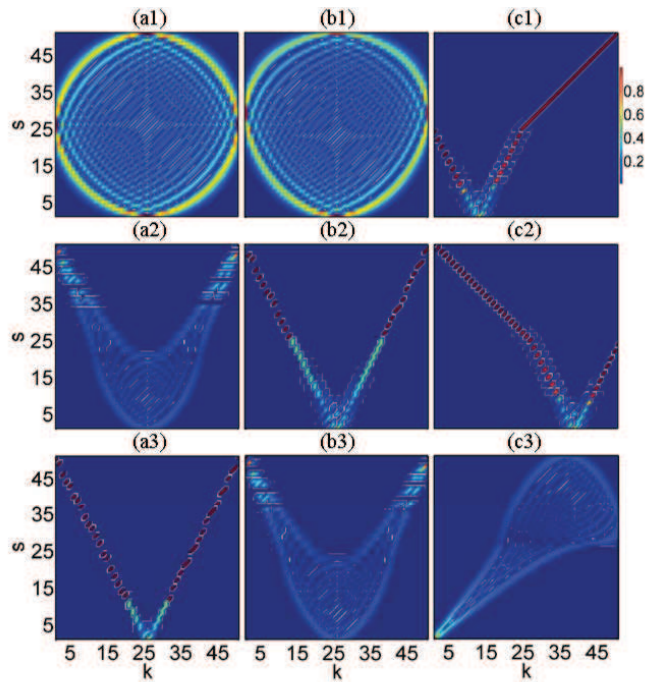}}
\caption{(color online) Density profiles $\left\vert c_{k}^{s}\right\vert
^{2}$, where $k=1,2,...,N+1$, and the ordinal number of the energy
eigenstate $s=1,2,...,N+1$. (a1) $U_{0}=0,U_{2}=0,\protect\mu =0$, (a2) $%
U_{0}=10,U_{2}=0,\protect\mu =0$, (a3) $U_{0}=500,U_{2}=0,\protect\mu =0$,
(b1) $U_{0}=0.5,U_{2}=0.02U_{0},\protect\mu =0$, (b2) $%
U_{0}=10,U_{2}=0.02U_{0},\protect\mu =0$, (b3) $U_{0}=500,U_{2}=0.02U_{0},%
\protect\mu =0$, (c1) $U_{0}=10,U_{2}=0.02U_{0},\protect\mu =5$, (c2) $%
U_{0}=10,U_{2}=0.02U_{0},\protect\mu =-5$, and (c3) $%
U_{0}=-10,U_{2}=0.02U_{0},\protect\mu =5$. Here $\protect\theta =0$ and $%
N=50 $.}
\end{figure}

In the meantime, we find that the density profiles of the stationary states
are significantly influenced by the pair tunneling and the tilt parameter of
the DW potential. For weak repulsive interaction as well as weak pair
tunneling, the density distributions are similar to those in figure 1(a1)
except for a further broading in the upper region [figure 1(b1)]. However,
for large repulsive interaction, pair tunneling exhibits distinct\ effect on
the density profiles, as shown in figures 1(b2) and 1(b3). Essentially, the
single-particle Josephson tunneling is suppressed for strong repulsion and
the pair tunneling gradually becomes dominant with continuously increasing
interactions. As mentioned above, the coupling constant $U_{3}=2.5U_{2}$ is
in direct proportion to the pair tunneling. In the case of $U_{0}=10$ and $%
U_{2}=0.02U_{0}$, the term of $2U_{3}\left( \hat{n}_{1}+\hat{n}_{2}-1\right)
/N^{2}$ in equation (\ref{ReducedHamiltonian}) actually suppresses the
interwell hoping. Thus the pair tunneling tends to make the eigenstates more
localized in comparison with the case of $U_{0}=10$ and $U_{2}=0$, which can
be seen in figure 1(b2). In addition, the effective single-atom Josephson
tunneling constant $(1/N-2U_{3}\left( N-1\right) /N^{2})$ can become a
negative value for sufficiently large $U_{3}$ (i.e., for sufficiently large
pair tunneling $U_{2}$ according to the relation $U_{3}=2.5U_{2}$).
Consequently, for the case of superstrong interaction $U_{0}=500$, the pair
tunneling $U_{2}=0.02U_{0}$ does not make the eigenstates further localized
but more extended instead [see figure 1(b3)]. The effect of the bias of the
double well on the density profiles is shown in figures 1(c1)-1(c3). When $%
\mu >0$ and $U_{0}>0$, the density profiles of the low-lying states display
a shift towards the left side and those of the high-lying states keep only
the right branch [figure 1(c1)] due to the symmetry breaking of the
left-right wells, while the case is reversed when $\mu <0$ and $U_{0}>0$
[figure 1(c2)]. Furthermore, the combining effect of attractive interaction
and pair tunneling as well as the bias of the DW potential is demonstrated
in figure 1(c3).

\begin{figure}[tbp]
\centerline{\includegraphics*[width=8.2cm]{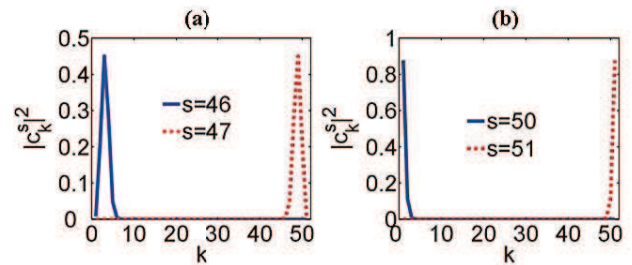}}
\caption{(color online) Density profiles $\left\vert c_{k}^{s}\right\vert
^{2}$ for typical degenerate eigenstate pairs in figure 1(a2). (a) The 46th
eigenstate (solid blue line) and the 47th eigenstate (dashed red line), and
(b) the 50th eigenstate (solid blue line) and the 51st eigenstate (dashed
red line). The distributions of these high-lying states are strongly
localized in the Fock space.}
\end{figure}

Figure 2 shows the density profiles $\left\vert c_{k}^{s}\right\vert ^{2}$
for the 46th, 47th, 50th, and 51st eigenstates in figure 1(a2), where the
former two states and the latter ones are two-fold degenerate, respectively.
Obviously, these high-lying excited states are strongly localized in the
Fock space. In the absence of pair tunneling, a dynamical quantum phase
transition from the Josephson regime to the self-trapped regime in a DW BEC
has been discussed recently in Ref. \cite{JuliaDiaz1}.

In figure 3, we present the lowest nine (ten) eigenvalues of the DW system
for repulsive (attractive) interatomic interaction as a function of the
ratio $r=U_{2}/[N-2U_{3}(N-1)]$ between the pair tunneling strength and the
single-particle tunneling strength, where $N=50$, $\mu =0$, and $\theta =0$.
It is shown that for strong repulsive interaction $U_{0}=100$ and zero pair
tunneling the lowest seven eigenstates are non-degenerate and pair
degeneracy occurs from the eighth eigenstate [figure 3(a)]. With the
increasing of $r$ (i.e., with the increasing of pair tunneling) below a
critical value $r_{c}=0.2$ the energy gap between two nearest-neighbor
non-degenerate eigenstates decreases remarkably. However, with the further
increase of $r$ the energy levels almost keep constant and exhibit a
characteristic of degeneracy or quasi-degeneracy. By contrast, for
attractive interaction the lowest ten eigenstates are pair degeneracy (or
quasi-degeneracy) irrespective of the pair tunneling strength, where the
relations between $E$ and $r$ are similar to those in figure 3(a).
Physically, the degeneracy or quasi-degeneracy of the energy levels are
resulted from the interatomic interactions (including the on-site two-body
interaction, the inter-well particle interaction, and the pair tunneling).
According to the definition of the tunneling ratio $r$, the pair tunneling
strength can be expressed by $U_{2}=N/[1/r+5(N-1)]$ because of $%
U_{3}=2.5U_{2}$. When $r>r_{c}$, the pair tunneling strength $U_{2}$
approaches an asymptotic value $N/5(N-1)$. Therefore with the further
increase of $r$ the eigenenergies of the system tend to keep constant, which
can be seen in the reduced Hamiltonian of equation (\ref{ReducedHamiltonian}%
). The present results indicate that the stability of the ground state (both
for repulsive and attractive interactions) is not affected basically by the
tunneling ratio $r$ as long as it is not less than a critical value.

\begin{figure}[tbp]
\centerline{\includegraphics*[width=6.6cm]{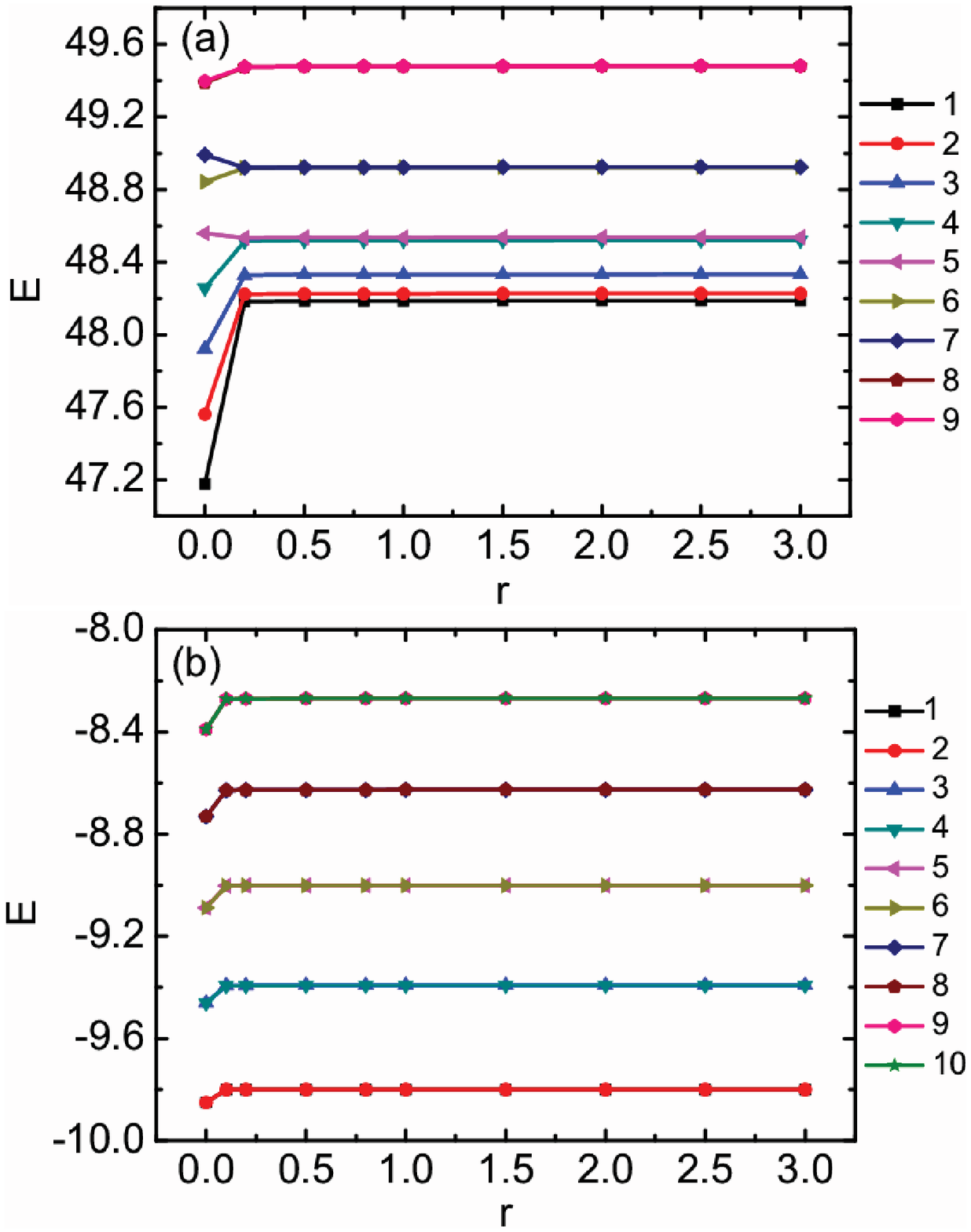}}
\caption{(color online) (a) The lowest nine energy levels (1--9) as a
function of the ratio $r$ between the pair tunneling strength and the
single-particle tunneling strength, where the relevant parameters are $%
U_{0}=100$, $N=50$, $\protect\mu =0$, and $\protect\theta =0$. (b) The
lowest ten energy levels (1--10) as a function of the ratio $r$. The
parameters are $U_{0}=-10$, $N=50$, $\protect\mu =0$, and $\protect\theta =0$%
.}
\end{figure}

Here we find that the Peierls phase $\theta $ does not influence the
stationary energy spectrums and the density profiles of the eigenvectors of
the DW system. This can be understood in terms of the expressions of the
Hamiltonian (\ref{StationaryHamiltonian}) and the Fock states (\ref%
{FockState}). If we make a gauge transformation $e^{i\theta }\hat{a}%
_{1}^{\dag }\rightarrow \hat{a}_{1}^{\dag }$, equation (\ref%
{StationaryHamiltonian}) reduces to the Hamiltonian of the DW system without
terms of Peierls phase.

\section{Quantum dynamics of the DW system with a maximal initial population
difference}

Next, we analyze the quantum dynamics of the DW system with a maximal
initial population difference $z(0)=1$ and $\left\vert \Psi (0)\right\rangle
=\left\vert N,0\right\rangle $. We numerically solve the coupled equations (%
\ref{C1Evolution})-(\ref{CN1Evolution}) using the fourth-order Runge-Kutta
method. The most important physical quantity characterizing the tunneling
dynamics and quantum fluctuations of the DW system is the population
difference. For the case of non-interacting limit and symmetric DW
potential, the system consists of $N$ independent particles and thus the
evolution of the population difference yields an obvious Rabi oscillation
\cite{Smerzi,Milburn,Ananikian,JuliaDiaz1}, which can be seen in figure
4(a1). In the presence of the bias of the double well, the population
difference exhibits a damped oscillation and finally approaches a large
steady value due to the incommensurate tunneling between the two wells
[figure 4(a2)]. In the typical Josephson regime (for instance, $U_{0}=1.2$, $%
U_{2}=0$, $\mu =0$), the evolution of the population difference always shows
damping oscillations followed by complex revivals as shown in figure 4(a3).
When the repulsive interatomic interaction, e.g., $U_{0}=4$, is larger than
a critical value, the population difference undergoes a damped oscillation
and then keeps a positive constant, which indicates that the system is in a
quantum self-trapping state [see figure 4(a4)]. For the case of repulsive
interactions, the quantum self-trapping effect is directly related to the
properties of the highest-lying state of the system Hamiltonian (\ref%
{ReducedHamiltonian}). As a matter of fact, when the initial quantum state
is prepared with strong repulsion $U_{0}$ above a critical value and with
large population difference, the system will remain trapped due to the large
overlap of the initial state with the highest excited state of the system.

\begin{figure}[tbp]
\centerline{\includegraphics*[width=8.5cm]{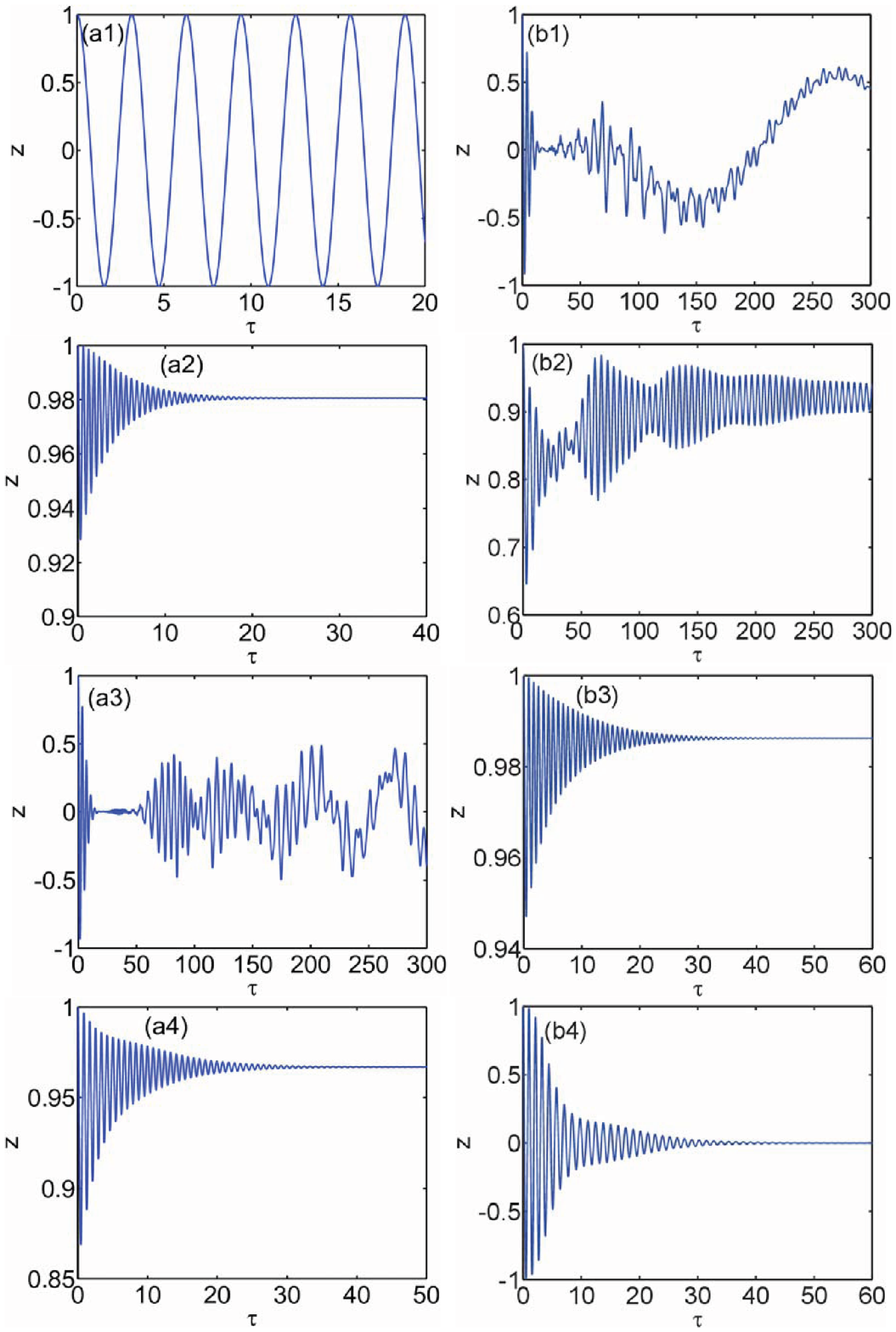}}
\caption{(color online) Time evolution of the population difference for a
quantum state of the DW system initially prepared with $\left\vert \Psi
(0)\right\rangle =\left\vert N,0\right\rangle $ and $z(0)=1$. (a1) $U_{0}=0$%
, $U_{2}=0$, $\protect\mu =0$, (a2) $U_{0}=0$, $U_{2}=0$, $\protect\mu =5$,
(a3) $U_{0}=1.2$, $U_{2}=0$, $\protect\mu =0$, (a4) $U_{0}=4$, $U_{2}=0$, $%
\protect\mu =0$, (b1) $U_{0}=1.2$, $U_{2}=0.02U_{0}$, $\protect\mu =0$, (b2)
$U_{0}=1.2$, $U_{2}=0.2U_{0}$, $\protect\mu =0$, (b3) $U_{0}=4$, $%
U_{2}=0.02U_{0}$, $\protect\mu =0$, and (b4) $U_{0}=4$, $U_{2}=0.2U_{0}$, $%
\protect\mu =0$. The Peierls phase is $\protect\theta =0$. Here the
horizontal ordinate $\protect\tau $ is in units of $\hbar /J$.}
\end{figure}

Not only does the pair tunneling influence remarkably the structure of the
stationary states but also the quantum dynamics of the DW system. In the
case of $U_{0}=1.2$, $U_{2}=0.02U_{0}$ and $\mu =0$, the
revival-collapse-revival evolution in the Josephson regime becomes a
modulated S-like oscillation as shown in figures 4(b1) and 4(a3), which
leads to a significant modification of the dynamics. When the pair tunneling
strength increases to $U_{2}=0.2U_{0}$ (strong pair tunneling), the
population difference oscillates anharmonically around an increasing time
averaged value of $\left\langle z(\tau )\right\rangle $ $\neq 0$. During the
long-time evolution of $z(\tau )$, the oscillation amplitude gradually
decreases and finally the population difference tends to keep a positive
constant value [figure 4(b2)]. Here the intriguing self-trapping phenomenon
is essentially resulted from the strong pair tunneling. For larger
interatomic interaction $U_{0}=4$, the medium strength pair tunneling $%
U_{2}=0.02U_{0}$ makes the oscillating population difference decay to a
higher constant value, which implies that weak and medium pair tunnelings
will significantly enhance the quantum self-trapping effect in a DW system
[figure 4(b3)]. By contrast, the strong pair tunneling $U_{2}=0.2U_{0}$ does
not strengthen the quantum self-trapping in the case of large repulsion, but
it destroys fully the self-trapping effect instead because of the final zero
population difference [figure 4(b4)]. From the foregoing analysis, we can
conclude that for the case of weak repulsion the strong pair tunneling
sustains a quantum self-trapping while for the case of large repulsion the
strong pair tunneling tends to destroy the quantum self-trapping effect.

\begin{figure}[tbp]
\centerline{\includegraphics*[width=8.5cm]{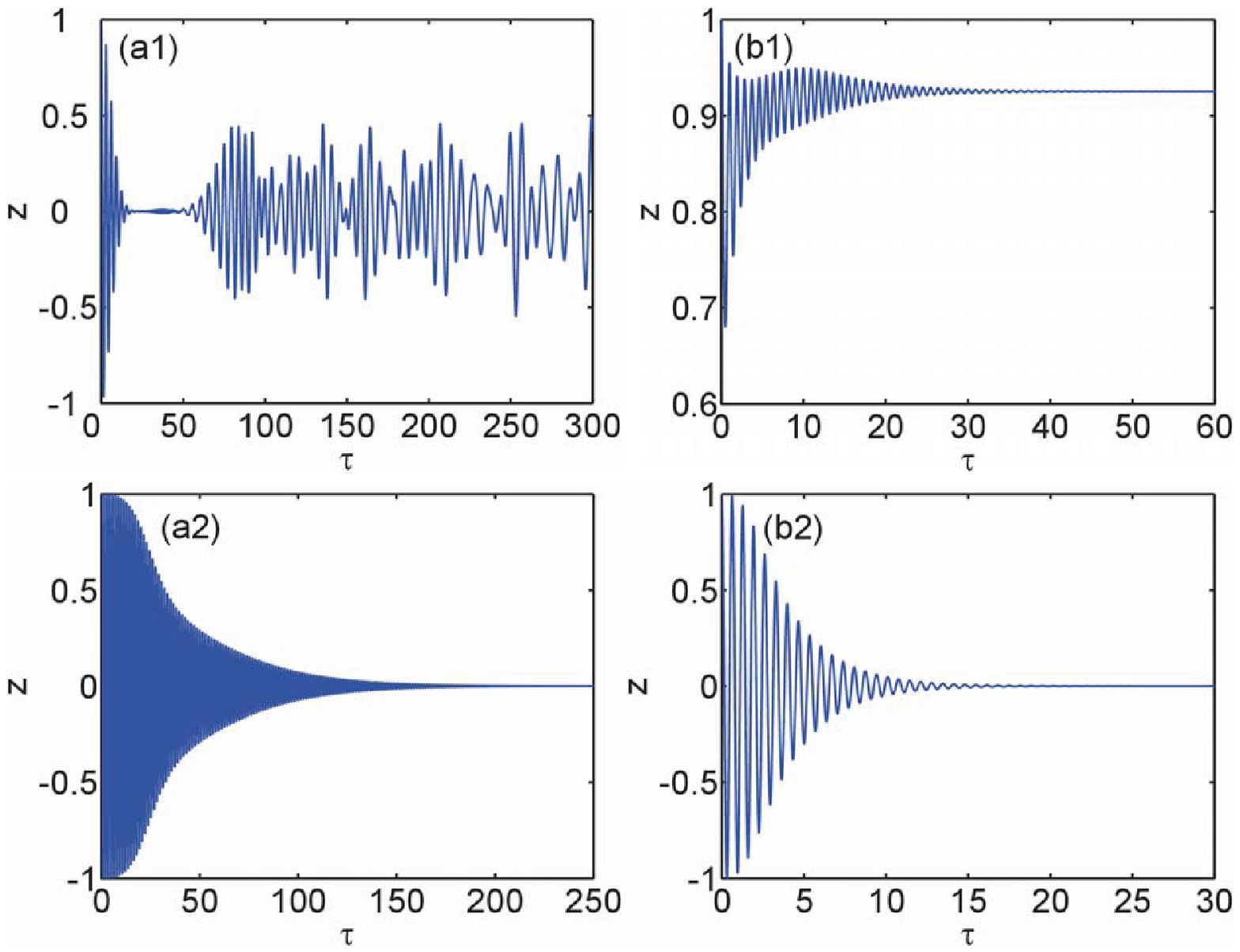}}
\caption{(color online) Time evolution of the population difference for a
quantum state of the DW system initially prepared with $\Psi (0)=\left\vert
N,0\right\rangle $ and $z(0)=1$. (a1) $U_{0}=-1.2$, $U_{2}=0.02U_{0}$, (a2) $%
U_{0}=-1.2$, $U_{2}=0.2U_{0}$, (b1) $U_{0}=-4$, $U_{2}=0.02U_{0}$, and (b2) $%
U_{0}=-4$, $U_{2}=0.2U_{0}$. The other parameters are $\protect\mu =0$ and $%
\protect\theta =0$. Here the horizontal ordinate $\protect\tau $ is in units
of $\hbar /J$.}
\end{figure}

Figure 5 displays the evolution of population difference for a DW system
with attractive interatomic interactions and pair tunneling, where $\mu =0$,
$\theta =0$, and the initial state of the system is $\left\vert \Psi
(0)\right\rangle =\left\vert N,0\right\rangle $ and $z(0)=1$. In the case of
$U_{0}=-1.2$ and $U_{2}=0.02U_{0}$ (weak attractive interaction and medium
strength pair tunneling), the population difference shows modulated
Josephson oscillations [figure 5(a1)], which is, to a certain extent,
similar to those in figures 4(a3) and 4(b1). For $U_{0}=-4$ and $%
U_{2}=0.02U_{0}$, the system exhibits obvious self-trapping behavior as
shown in figure 5(b1). In particular, we find that strong pair tunneling
thoroughly eliminates the self-trapping phenomenon in the DW system for both
weak and strong attractive interactions, where the ultimate population
difference remains zero [figures 5(a2) and 5(b2)]. Note that the shade
region in figure 5(a2) represents rapid oscillations of the population
difference.

\section{Quantum quench dynamics of the DW system in the presence of pair
tunneling}

Now we discuss the quantum quench dynamics of the DW system so that we can
explore the effect of the Peierls phase on the dynamic properties of the
system. We say that the system is quenched if the Peierls phase $\theta $ is
changed suddenly at a certain moment from one value to another. For
simplicity, suppose that at the initial time $\tau =0$ the DW system is
quenched by changing the value of $\theta $ from zero to a nonzero constant
value and the initial state $\left\vert \Psi (0)\right\rangle $ is the
ground state of the DW system with $\theta =0$. In order to ensure the
accuracy and reliability, we test in advance the dynamical evolution of the
ground state of the DW system before computing the quench dynamics of the
system for each given set of parameters. The reasonable time-dependent
population difference concerning the dynamics of the ground state of the
system should be a perfect straight line, which has been verified in our
simulations.

Figure 6 shows the quench dynamics of the DW system for repulsive
interatomic interactions, where $N=50$, $\mu =0$, and $U_{2}/U_{0}=0.002$.
For $0<\theta <\pi $, the larger the value of $\theta $ is, the larger the
maximum amplitude of the oscillation becomes, and at the same time the
shorter the quasi-period of collapse and revival gets. The case is reversed
for $\pi <\theta <2\pi $ (not shown here). In addition, for the same Peierls
phase $\theta $, the stronger the repulsive interaction is, the lower the
maximum amplitude of the oscillation is. This feature can be understood
because the large interatomic interaction is equivalent to a high potential
barrier which weakens the particle tunneling and tends to make the atoms
localized. We show that for repulsive interactions and $\theta =\pi $ there
exists no oscillation behavior during the dynamic evolution of the DW system
and the population difference keeps unchanged, i.e., $z(\tau )=0$. The main
reason is that the ground state of the system with $\theta =0$ (i.e, the
initial state) in the case of repulsive interatomic interactions is always
symmetric and thus a $\pi $ Peierls phase does not influence the
single-particle tunneling and the pair tunneling as shown in equations (\ref%
{StationaryHamiltonian}) and (\ref{ReducedHamiltonian}).

\begin{figure}[tbp]
\centerline{\includegraphics*[width=8.5cm]{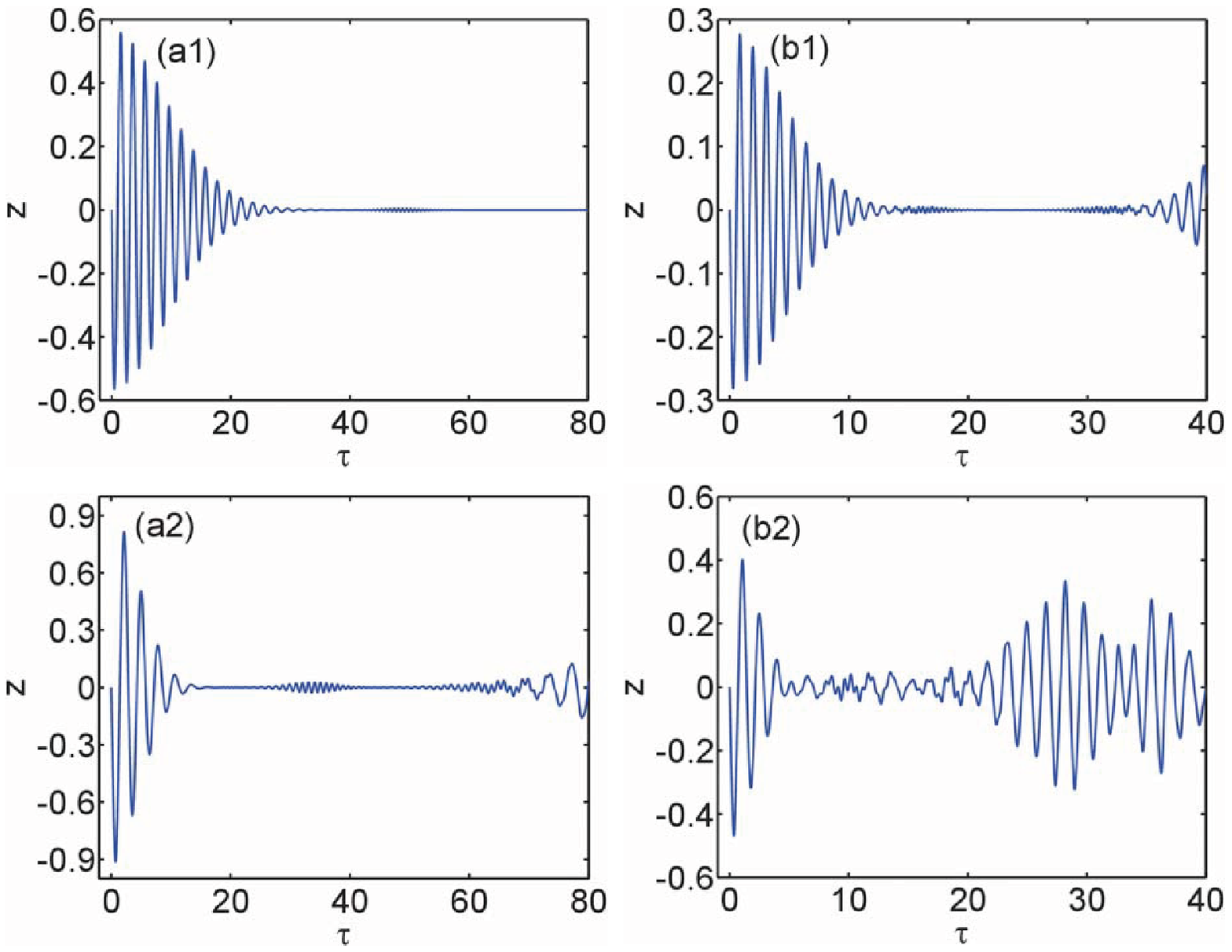}}
\caption{(color online) Quantum quench dynamics of the system for repulsive
interatomic interactions by suddenly changing the Peierls phase of the DW
potential, where the initial state is the ground state of the DW system with
$\protect\theta =0$. (a1) $U_{0}=2$, $\protect\theta =\protect\pi /3$, (a2) $%
U_{0}=2$, $\protect\theta =2\protect\pi /3$, (b1) $U_{0}=10$, $\protect%
\theta =\protect\pi /3$, and (b2) $U_{0}=10$, $\protect\theta =2\protect\pi %
/3$. The other parameters are $N=50$, $U_{2}/U_{0}=0.002 $, and $\protect\mu %
=0$. Here the horizontal ordinate $\protect\tau $ is in units of $\hbar /J$.}
\end{figure}

In the case of attractive interatomic interactions, the quantum quench
dynamics of the system becomes more complicated. For weak attractive
interaction $U_{0}=-2$ (Josephson oscillation regime), the oscillation is
somewhat erratic as shown in figures 7(a1) and 7(a2). Physically, particle
number fluctuations can automatically admix the excited states of the system
due to the many-body interactions. Thus the quench caused by the sudden
change of the Peierls phase generates various elementary excitations with
different frequencies. The similar feature is also found in the case of
few-bosons dynamics in a double well \cite{Zollner}. With the time
evolution, the population difference finally restores to its initial value $%
z=0$. From figures 7(a1) and 7(a2), one can see that when the Peierls phase $%
\theta $ ($0<\theta <\pi $) get larger the maximum value of the oscillation
amplitudes becomes larger, which is similar to the case of repulsive
interactions. In addition, we find that the population difference keeps zero
in the case of $\theta =\pi $ as shown in figure 7(a3). This point can be
understood because for the weak attractive interaction of $U_{0}=-2$ the
ground state of the DW system is still symmetric, which is similar to the
case of repulsive interactions. For strong attractive interaction and $%
\theta =0$, the ground sate of the DW system is a quantum self-trapping
state with a large population difference, i.e., the atoms are self trapped
in a single trap (e.g., the right trap) of the DW potential. Once the
Hamiltonian is quenched due to the sudden change of the Peierls phase $%
\theta $, the oscillation gradually decays and the population difference
eventually resumes its original value $z=-0.991$, which can be seen in
figures 7(b1)-7(b3). In the case of $0<\theta \leq \pi $, a larger Peierls
phase $\theta $ implies a higher oscillation amplitude. We expect that the
effect of the Peierls phase on the quench dynamics of the DW system can be
observed and tested in the future experiments.

\begin{figure}[tbp]
\centerline{\includegraphics*[width=8.5cm]{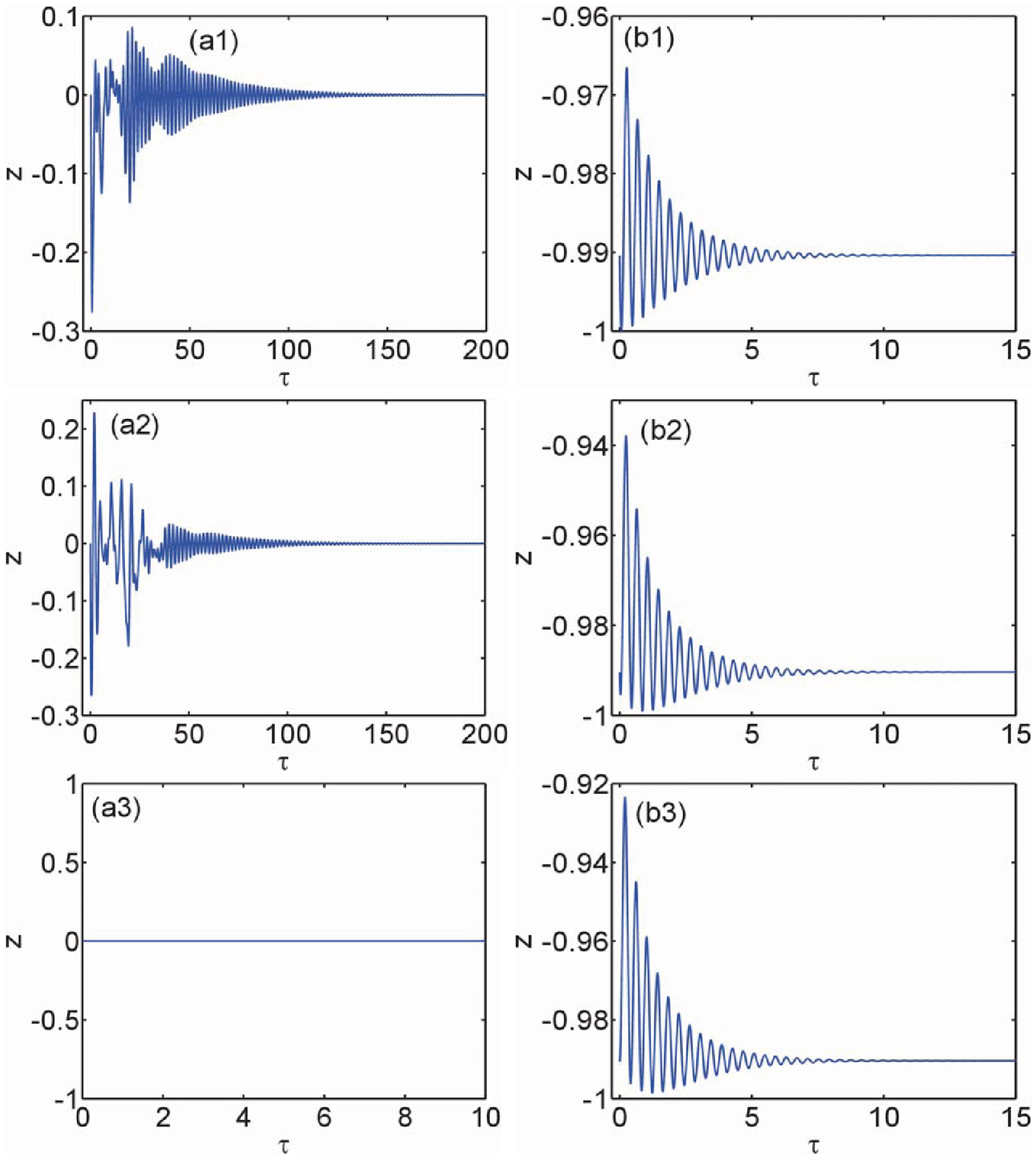}}
\caption{(color online) Quantum quench dynamics of the system for attractive
interatomic interactions by suddenly changing the Peierls phase of the DW
potential, where the initial state is the ground state of the DW system with
$\protect\theta =0$. (a1) $U_{0}=-2$, $\protect\theta =\protect\pi /3$, (a2)
$U_{0}=-2$, $\protect\theta =2\protect\pi /3$, (a3) $U_{0}=-2$, $\protect%
\theta =\protect\pi $, (b1) $U_{0}=-8$, $\protect\theta =\protect\pi /3$,
(b2) $U_{0}=-8$, $\protect\theta =2\protect\pi /3$, and (b3) $U_{0}=-8$, $%
\protect\theta =\protect\pi $. The other parameters are $N=50$, $%
U_{2}/U_{0}=0.002$, and $\protect\mu =0$. Here the horizontal ordinate $%
\protect\tau $ is in units of $\hbar /J$.}
\end{figure}

\section{Conclusion}

In summary, we have applied a full quantum-mechanical procedure to
investigate the properties of the stationary states of BECs in a DW
potential and the quantum dynamics of the system with a maximum initial
population difference. At the same time, the quantum quench dynamics of the
system is studied by suddenly changing the Peierls phase of the DW
potential. We show that the pair tunneling influences significantly the
density profiles and energy spectrums of the stationary states especially
for the cases of medium or strong interatomic interactions. With the
increase of the ratio between the pair tunneling strength and the
sing-particle tunneling strength, the eigenstates of the Hamiltonian become
pair degenerate and the corresponding energy levels ultimately keep almost
constant. In addition, it is shown that in the case of weak repulsion the
strong pair tunneling sustains a quantum self-trapping for the quantum
dynamics of the DW system with a maximum initial population difference while
in the case of large repulsion the strong pair tunneling tends to destroy
the self-trapping effect. By contrast, for the case of attractive
interactions the strong pair tunneling tends to eliminate the Josephson
oscillations and the quantum self-trapping irrespective of the interaction
strength. Moreover, when the DW system is quenched, the population
difference shows evident oscillation behaviors for most cases. The maximum
amplitude of the oscillation is directly relevant to the value of the
Peierls phase. These properties are remarkably different from those in a
conventional DW BEC with no pair tunneling, which allows to be observed and
tested in the future experiments.

\begin{acknowledgments}
We thank Biao Wu, Ming Gong, and Chunlei Qu for helpful discussions and
comments. This work is supported by NSFC (Grants No. 11475144 and No.
11304270), NSF of Hebei Province (Grant No. A2015203037), and Ph.D.
foundation of Yanshan University (Grant No. B846).
\end{acknowledgments}

\end{document}